\documentstyle[12pt]{article}
\begin{document}
\thispagestyle{empty}
\setcounter{page}{0}
\renewcommand{\thefootnote}{\fnsymbol{footnote}}
\hskip 10cm {\sl HUB-EP-95/24}
\vskip.0pt
\hskip 10cm hep-th/9510
\begin{center}
{\Large \bf On a possibility to construct gauge
} \\[1ex]
{\Large\bf invariant quantum formulation
}\\[1ex]
{\Large \bf for non-gauge classical theory
} \\[11ex]
 {\large I. L. Buchbinder
\footnote{Supported by Deutsche Forschungsgemeinschaft
under contract DFG- 436 RUS 113}
}\\
 {\it
 Humboldt-Universit\"at zu Berlin\\
  Institut f\"ur Physik\\
  Theorie der Elementarteilchen\\
D-10115 Berlin, Germany\\
and\\
Department of Theoretical Physics\\
Tomsk State Pedagogical University\\
Tomsk 634041, Russia\\}

\vskip.2cm
 {\large V. D. Pershin, G. B. Toder}\\
 {\it
 Department of Theoretical Physics\\
Tomsk State University\\
Tomsk 634050, Russia\medskip\\
\
}
\vskip.5cm
 {\large \bf Abstract}
\end{center}
\begin{quotation}
A non-gauge dynamical system depending on parameters is considered.
It is shown that these parameters can have such values that
corresponding canonically quantized theory will be gauge invariant.
The equations allowing to find these values of parameters are derived. The
prescription under consideration is applied to obtaining the equation
of motion for tachyon background field in closed bosonic string
theory.
\end{quotation} \clearpage
\renewcommand{\thefootnote}{\protect\arabic{footnote}}

\newpage
\section{Introduction}
Procedure of canonical quantization is a natural and correct
base for construction of consistent quantum theory and it is
applied as a ground for any specific quantization methods.
Formulation of new fundamental models of quantum
field theory leads, as a rule, to necessity of
studying new aspects of canonical quantization. At
present the most general approach to canonical quantization of
arbitrary systems is BFV-method~\cite{FrVil_75}-\cite{BatFr_88} (see
also~\cite{Hen_85,HenTei_92}) which includes all the previous achievements
of other quantization methods.

In this paper we would like to discuss a new
interesting aspect of canonical
quantization arising from the string theory in
background fields~\cite{Lov_84}-\cite{Sen_85 2}. In this theory
one considers a string interacting with massless background fields
and introduces the Weyl invariance principle according to which the
renormalized trace of energy-momentum tensor should vanish.
General structure of the trace was studied in
refs.~\cite{Tse_87,Os_87}.

The consideration carried out in refs.~\cite{Lov_84}-\cite{Os_87}
in schematic form looks as follows. Let us consider for simplicity
the bosonic string theory only. In this case set of dynamical
variables consists of string coordinates $X^{\mu}$ and  components
of two-dimensional world sheet metric $g_{ab}(\tau,\sigma)$.
Classical lagrangian includes the Fradkin-Tseytlin
term~\cite{FrTse_85 1,FrTse_85 2} describing  string
interaction with dilaton field. Presence of this term spoils
Weyl invariance of  the classical theory and
trace of classical energy-momentum tensor does not vanish
identically. Neverthereless, effective action depending on
external fields $g_{ab}$ is considered and condition of its
independence on conformal factor of two-dimensional metric is
imposed. This condition leads to equations of motion for background
fields and their explicit form can be found perturbatively.

It is important that there is no Weyl
anomaly and its cancellation in such a theory.
The initial classical theory is not Weyl
invariant and therefore there is no sense to say about anomalies.
An analogous situation also arises in string theory interacting with
tachyon field or with massive background field of higher
levels~\cite{BuchKry_95,FubRon_88} where
classical theory is not Weyl invariant but at the quantum level
Weyl invariance takes place under equations of motion for
background fields. Note that this approach
suggests calculation of functional integral in a strictly specified
way. Namely, one should first calculate the integral over string
coordinates and then demand effective action to be Weyl invariant.
Thereafter the dependence on two-dimensional metric will be reduced
to dependence on a finite number of parameters defining the world
sheet topology.\footnote{In the recent paper~\cite{BuchSh_95} an
attempt to treat the integral over string coordinates and world sheet
metric without taking into account any special order of integration
was undertaken.}.

From general point of view the above situation can be formulated as
follows. Consider a non-gauge classical theory action which
depends on some parameters, in string theory their role
being played by background fields. If the given parameters
are constrained by special restrictions then the effective action
satisfies identities defining quantum gauge invariant theory.
The problem is how to describe such a situation in terms of canonical
quantization.

It is well known that any
quantum gauge theory is characterized by the operators of first
class constraints and their algebra is generated by the
nilpotency condition of canonical BRST-charge
$\Omega$~\cite{FrVil_75}-\cite{BatFr_88}. The  constraints operators
are usually constructed on the base of classical theory  constraints.
However, if classical theory is not gauge-invariant it has no first
class constraints at all and the operator $\Omega$ can not be
constructed. Thus, from the point of view of standard canonical
quantization the situation when gauge quantum theory corresponds to
non-gauge classic theory looks very strange.  At the same time,
we believe that any adequate quantization procedure should be
in agreement with canonical quantization\footnote{In
particular, for bosonic string interacting with background
gravitational field this agreement was established~\cite{BuchMi_95}.
However, in that case we had no the problem under consideration.}.

This paper is devoted to a possible way of solving this problem.
We introduce some prescription allowing to construct the operator
$\Omega$ starting from non-gauge classical theory which depends on
some parameters and show that this operator $\Omega$ will be
nilpotent only under special equations for the parameters. Then we
derive  equation of motion for tachyon background
field in closed bosonic string theory using the introduced
recipe and demonstrate its efficiency.

\section{General Formulation}

Let us consider a quantum system with the Hamiltonian of the
following form:
\begin{equation}
H=H_{0}(a)+\lambda^{\alpha}T_{\alpha}(a).
\end{equation}
Here $\lambda^{\alpha}$, $a\equiv a_{I}$
are classical parameters of the theory,
$H_{0}(a)\equiv H_{0}(q^{i},p_{i}|a)$,
$T_{\alpha}(a)\equiv T_{\alpha}(q^{i},p_{i}|a)$;
$(q^{i},p_{i})$ are operators of canonically conjugated dynamical
variables. Since $\lambda^{\alpha}$ are given parameters we cannot
consider $T_{\alpha}(a)$ as constraints.

We suggest that operators $T_{\alpha}(a)$ can be presented in
the form
\begin{equation}
T_{\alpha}(a)=T^{(0)}_{\alpha}(a)+T^{(1)}_{\alpha}(a),
\end{equation}
where
\begin{eqnarray}
[T^{(0)}_{\alpha}(a),T^{(0)}_{\beta}(a)]& = &i\hbar
T^{(0)}_{\gamma}(a)U^{\gamma}
_{\alpha\beta}(a),\nonumber \\ {}
[H_{(0)}(a),T^{(0)}_{\alpha}(a)]& = &i\hbar T^{(0)}_{\gamma}(a)V^{\gamma}
_{\alpha}(a).
\end{eqnarray}
However
\begin{eqnarray}
[T_{\alpha}(a),T_{\beta}(a)]& = &i\hbar T_{\gamma}(a)U^{\gamma}
_{\alpha\beta}(a)+i\hbar A_{\alpha\beta}(a)\nonumber \\ {}
[H_{(0)}(a),T_{\alpha}(a)]& = &i\hbar T_{\gamma}(a)V^{\gamma}_{\alpha}(a)
+i\hbar A_{\alpha}(a)
\end{eqnarray}
and the operators $A_{\alpha\beta}(a)$, $A_{\alpha}(a)$ do not vanish
in the classical limit. It means that corresponding classical
theory is not gauge invariant.

Let us introduce  operators $\Omega$ and $H$ formally following
the BFV-method as if $T_{\alpha}$ corresponded to
first class constraints:
\begin{eqnarray}
\Omega & = &
c^{\alpha}T_{\alpha}(a)-{1\over2}U^{\gamma}_{\alpha\beta}(a)
O({\cal P_{\gamma}} c^{\alpha} c^{\beta})\nonumber \\
H & = & H_{(0)}(a)+V^{\gamma}_{\alpha}(a)O({\cal P_{\gamma}}
c^{\alpha})
\end{eqnarray}
where $O$ denotes some suitable ordering of ghost
operators $\cal P_{\gamma}$, $c^{\alpha}$.

It is not difficult to show that
\begin{eqnarray}
\Omega^{2} & = & {1\over2}([T_{\alpha}(a),T_{\beta}(a)]-
i\hbar T_{\gamma}(a)U^{\gamma}_{\alpha\beta}(a)+i\hbar G_{\alpha\beta}(a))
c^{\alpha}c^{\beta}
\nonumber \\
\frac{d\Omega}{dt}& = &c^{\alpha}\frac{\partial T_{\alpha}(a)}{\partial t}-
{1\over2}\frac{\partial U^{\gamma}_{\alpha\beta}(a)}{\partial t}
O({\cal P_{\gamma}}c^{\alpha}c^{\beta})
\\
 & - &(i\hbar)^{-1}
([H_{(0)}(a),T_{\alpha}(a)]-i\hbar T_{\gamma}(a)V^{\gamma}_{\alpha}(a)
+i\hbar G_{\alpha}(a)c^{\alpha})
\nonumber
\end{eqnarray}
where $G_{\alpha\beta}(a)$ and $G_{\alpha}(a)$ are  possible
ghost contributions. Here we took into account the possible
explicit dependence of the operators $T_{\alpha}(a)$ on
time\footnote{Formulation of BFV-procedure for theories
explicitly depending  on time was given in ref.~\cite{BatLya_90}.}.
The eqs.(4,6) lead to
\begin{eqnarray}
\Omega^{2} & = &
{1\over2}i\hbar E_{\alpha\beta}(a)c^{\alpha}c^{\beta} \nonumber \\
\frac{d\Omega}{dt}& = &\left( \frac{\partial T_{\alpha}(a)}{\partial t}-
E_{\alpha}(a)\right) c^{\alpha}-{1\over2}
\frac{\partial U^{\gamma}_{\alpha\beta}(a)}{\partial t}
O({\cal P_{\gamma}}c^{\alpha}c^{\beta})
\end{eqnarray}
where
\begin{eqnarray}
E_{\alpha\beta}(a)& = &A_{\alpha\beta}(a)+G_{\alpha\beta}(a)
\nonumber \\
E_{\alpha}(a)& = &A_{\alpha}(a)+G_{\alpha}(a)
\end{eqnarray}

We suppose that the operators
$E_{\alpha\beta}(a)$ and
${\partial T_{\alpha}(a)}/{\partial t}-E_{\alpha}(a)$
can be presented as linear combinations of an irreducible set of
operators $O_{M}\equiv O_{M}(q^{i},p_{i})$:
\begin{eqnarray}
E_{\alpha\beta}(a)& = &E^{M}_{\alpha\beta}(a)O_{M}
\nonumber \\
\frac{\partial T_{\alpha}(a)}{\partial t}-E_{\alpha}(a)& = &
E^{M}_{\alpha}(a)O_{M}
\end{eqnarray}
where  $E^{M}_{\alpha\beta}(a)$, $E^{M}_{\alpha}(a)$ are
$c$-valued functions of the parameters $a$. Then one gets
\begin{eqnarray}
\Omega^{2}& = &{1\over2}ihE^{M}_{\alpha\beta}(a)O_{M}
c^{\alpha}c^{\beta}
\nonumber \\{}
\frac{d\Omega}{dt}& = &E^{M}_{\alpha}(a)O_{M}c^{\alpha}
-{1\over2}\frac{\partial U^{\gamma}_{\alpha\beta}(a)}{\partial t}
O({\cal P_{\gamma}}c^{\alpha}c^{\beta})
\end{eqnarray}

We see that in general case $\Omega^{2}\neq 0$,
${d\Omega}/{dt}\neq 0$. But the theory can not
be called an anomalous one since classical symmetries are absent
and it does not make sense to say about their violation at quantum
level\footnote{The general discussion of anomalies in the
BFV-approach can be found in ref.~\cite{Mar_87}.}.

Let us suppose that there exists a solution of the following system
of equations for the parameters $a$:
\begin{eqnarray}
E^{M}_{\alpha\beta}(a)& = &0\nonumber \\
E^{M}_{\alpha}(a)& = &0
\end{eqnarray}
We also suppose that the equation
$$\frac{\partial U^{\gamma}_{\alpha\beta}(a)}{\partial t}=0$$
is fulfilled. Let us denote corresponding solution for the
parameters as $a^{(0)}\equiv a^{(0)}_{I}$. Then
\begin{eqnarray}
\Omega^{2}\biggr|_{\displaystyle a=a^{(0)}}& = &0
\nonumber \\
\frac{d\Omega}{dt}\biggr|_{\displaystyle a=a^{(0)}}& = &0
\end{eqnarray}

Thus, when parameters $a$ in eqs.(1,2,4,5) take the values
$a^{(0)}$ the equations $\Omega^{2}=0$,
${d\Omega}/{dt}=0$ defining a quantum gauge theory should
take place. As a result we get the prescription allowing
to construct a gauge invariant quantum formulation
for non-gauge classical theory depending on parameters\footnote{In
specific theories the equations (11) may have no solutions and
corresponding gauge-invariant quantum formulation may not exist at
all.  Also a part of equations (11) may be fulfilled
identically. An example of theory  where all the equations (11) are
fulfilled identically   was given in the paper~\cite{FubRon_88}.}.

\section{Application to String Coupled to Tachyon Background Field}

To illustrate efficiency of the above recipe we consider a
derivation of linear equation of motion for the tachyon field in
closed bosonic string theory.\footnote{It is well known that to obtain
the tachyon equation of motion consistent with
structure of string amplitudes one should use a nonperturbative
consideration (see e.g.~\cite{DasSat_86,Itoi_87}).
We restrict ourselves to linear approximation since our
purpose here is just to illustrate that the above prescription really
works.}
The theory is described by the following action:
\begin{equation}
S=-(2\pi\alpha^\prime)^{-1} \int\,d^{2}\sigma\,
 \sqrt{-g} \{ g_{ab}\partial_{a} X^{\mu} \partial_{b}
 X^{\nu} \eta_{\mu\nu} +Q(X)\}
\end{equation}
Here $\sigma^{a}\equiv (\tau,\sigma)$; $a,b=0,1$;
$\mu,\nu=0,1,\ldots, D-1$; $\eta_{\mu\nu}$ is  Minkowski
metric of $D$-dimensional background space-time, $Q(X)$ is
tachyon background field.

It is easy to show that
\begin{equation}
g_{ab} \frac{\delta S}{\delta g_{ab}}=-
(2\pi\alpha^\prime)^{-1}\sqrt{-g}Q(X)
\end{equation}
Therefore, if the metric $g_{ab}$ was a dynamical variable
then the corresponding classical equations of motion  would lead to
$Q(X)=0$. The analogous situation  takes place for string
theory interacting with either the dilaton field~\cite{BuchFr_91} or
the massive higher spin background
fields~\cite{BuchKry_95,FubRon_88}. To fulfill classical equations of
motion for two-dimensional metric one should set the dilaton field
to be constant and all the higher massive background fields equal to
zero. In order to avoid this situation we have to conclude that
components of the metric $g_{ab}$ should be treated as external
fields. Such a conclusion is consistent with general ideology
accepted in string theory interacting with background
fields~\cite{Lov_84} - \cite{BuchKry_95} where functional
integral is calculated only over variables $X^{\mu}$ and metric
components $g_{ab}$ are considered as external fields.

Let us parametrize the metric $g_{ab}$ as follows~\cite{BuchFr_91}
\begin{eqnarray}
g_{ab}& = &e^{\gamma}\left(\begin{array}{cc}
\lambda^{2}_{1}-\lambda^{2}_{0} & \lambda_{1} \\
\lambda_{1}  & 1 \end{array}\right)
\end{eqnarray}
It is easy to show that Hamiltonian of the theory (13) has the form
\begin{equation}
H=\int\,d\sigma\,(\lambda_{0}T_{0}+\lambda_{1}T_{1})
\end{equation}
where
\begin{eqnarray}
T_{0}& = &{1\over2}\left( (2\pi\alpha^\prime)P_{\mu}P^{\mu}+
(2\pi\alpha^\prime)^{-1}X^\prime_{\mu}X^{\prime\mu}\right) +
(2\pi\alpha^\prime)^{-1}e^{\gamma}Q(X)
\nonumber \\
T_{1}& = &P_{\mu}X^{\prime\mu}
\end{eqnarray}
Here  $P_{\mu}$ are
momenta canonically conjugated to coordinates $X^{\mu}$ and
$X^{\prime\mu}=\partial_{\sigma}X^{\mu}$.
$\lambda_{0}$, $\lambda_{1}$ are external fields and  role of the
parameters $a$ is played by the tachyon field $Q(X)$. $\gamma$ is
an external field also. The eq.(16) corresponds to the case
$H_{0}=0$.

Let us introduce the functions
\begin{eqnarray}
L & = & {1\over2}(T_{0}-T_{1})
 = {1\over4}(2\pi\alpha^\prime)^{-1}
\left( (2\pi\alpha^\prime)P-X^{\prime}\right)^{2}
\nonumber \\
 & + & (4\pi\alpha^\prime)^{-1}e^{\gamma}Q(X)
\nonumber \\
\bar{L} & = & {1\over2}(T_{0}+T_{1}) = {1\over4}(2\pi\alpha^\prime)^{-1}
\left( (2\pi\alpha^\prime)P_+X^{\prime}\right)^{2}
\nonumber \\
 & + & (4\pi\alpha^\prime)^{-1}e^{\gamma}Q(X)
\end{eqnarray}

It is evident that
\begin{equation}
L=L^{(0)}+L^{(1)},
\bar{L}=\bar{L}^{(0)}+\bar{L}^{(1)}
\end{equation}
where $L^{(0)}$ and $\bar{L}^{(0)}$ represent standard constraints of
the free string theory satisfying Virasoro algebra and
\begin{equation}
L^{(1)}=\bar{L}^{(1)}=(4\pi\alpha^\prime)^{-1}e^{\gamma}Q(X)
\end{equation}
We pay attention that the functions $L$ and $\bar{L}$ are not
constraints and they do not satisfy any algebra.\footnote{If we
considered $\lambda_{0}$, $\lambda_{1}$ and $\gamma$ as dynamical
variables then, of course, $T_{0}$  and $T_{1}$ would be constraints,
$Q(X)=0$ and the corresponding algebra of first class constraints
would have standard form of Virasoro algebra. We study another
situation when $Q(X)\neq0$. In this case $\lambda_{0}$,
$\lambda_{1}$ and $\gamma$  are  external fields, $T_{0}$  and
$T_{1}$ are not constraints and their Poisson brackets
are not expressed in terms of $T_{0}$  and $T_{1}$.}

Let us pass now to quantum theory. We choose conformal gauge
for the external fields $\lambda_{0}=1$, $\lambda_{1}=0$. In order to
obtain linear equation for $Q$ it is sufficient to suppose that
$X^{\mu}$ satisfy free string equation of motion $\Box X^{\mu}=0$.
It means that we can introduce  zero and oscillating string modes
operators by the standard way.

Let $:L(\tau, \sigma):$,
$:\bar{L}(\tau, \sigma):$ are the operators corresponding
to the classical functions (18) ordered according to the
prescription defined in the Appendix A. We introduce the operators
\begin{eqnarray}
:L_n (\tau): & = & \int^{2\pi}_0
\,d\sigma\, e^{-\imath n\sigma}:L(\tau, \sigma):
\nonumber \\
:\bar{L}_n (\tau): & = & \int^{2\pi}_0 \,d\sigma\,
e^{\imath n\sigma}:\bar{L}(\tau, \sigma):
\end{eqnarray}

Our main aim is to construct the eqs.(11) in the case under
consideration. Let us start with the first of these equations
and find the operators corresponding to
$A_{\alpha\beta}$ (4) and $G_{\alpha\beta}$ (6). The eqs.(3)
show that in order to find $A_{\alpha\beta}$ it is necessary to
calculate the commutators $[:L_n: , :L_m:]$,
$[:\bar{L}_n:,:\bar{L}_m:]$, $[:L_n:,:\bar{L}_m:]$ where $:L_n:$ and
$:\bar{L}_m:$ are given by (21).

To compute the above commutators we introduce  symbols
$L_n$ and $\bar{L}_m$ of operators $:L_n:$ and $:\bar{L}_m:$
associated with the given ordering prescription.

It is well known that calculation of the commutators
can be reduced to calculation of so called $*$-commutators for the
corresponding symbols~\cite{Ber_83,Vas_76}. An example of such
a calculation in string theory is given in ref.~\cite{BuchFr_91}.

Let us denote the $*$-commutator of the symbols under
consideration as $[L_n , L_m ]_* $, $[\bar{L}_n , \bar{L}_m ]_* $, $[L_n ,
\bar{L}_m ]_* $. Calculating these $*$-commutators we obtain some
integrals (see Appendix B) which
unfortunately are ill defined  and we
should use some regularization procedure. Since we are
constructing the calculating scheme within the canonical
approach we have no possibility to apply standard
regularization procedures accepted for regularization of Feynman
integrals in  covariant approaches. To regularize the integrals
under consideration we have used the method given in Appendix B.

The result of calculations of $*$-commutators can be written as
follows\footnote{We pay attention that regularized
$*$-commutators do not depend on the regularization parameter. It
means that the composite operator $\Omega^2$ is finite
automatically and it does not need renormalization in the
theory under consideration.}
\begin{eqnarray}
[L_n , L_m ]_* & = & \hbar (n-m)L_{n+m} +
\hbar^2\alpha_0 (n-m)\delta_{n,-m} + {D\over 12}\hbar^2 (n^3 -
n)\delta_{n,-m}
\nonumber \\
 & - & (4\pi\alpha^\prime )^{-1}\hbar (n-m)\int^{2\pi}_0 \,
d\sigma\, e^{-\imath(n+m)\sigma} e^{\gamma (\tau , \sigma)}
\left( 1+\alpha^\prime \hbar\Box  /4\right) Q(X)
\nonumber \\
{}[\bar{L}_n , \bar{L}_m ]_* & = & \hbar (n-m)\bar{L}_{n+m} +
\hbar^2 \beta_{0}(n-m)\delta_{n,-m} + {D\over 12}\hbar^2 (n^3 -
n)\delta_{n,-m}
\nonumber \\
 & - & (4\pi\alpha^\prime )^{-1}\hbar (n-m)\int^{2\pi}_0 \,
d\sigma\, e^{\imath(n+m)\sigma} e^{\gamma (\tau , \sigma)}
\left( 1+\alpha^\prime \hbar\Box /4\right) Q(X)
\nonumber \\
{}[L_n , \bar{L}_m ]_* & = & -(4\pi\alpha^\prime )^{-1}\hbar
(n-m)\int^{2\pi}_0 \, d\sigma\, e^{\imath(m-n)\sigma} e^{\gamma
(\tau , \sigma)} \left( 1+\alpha^\prime \hbar\Box /4\right)Q(X)
\nonumber \\
& - & (4\pi\alpha^\prime )^{-1} i\hbar
 \int^{2\pi}_0 \, d\sigma\, e^{\imath(m-n)\sigma} \left(
e^{\gamma (\tau , \sigma)}\right)^\prime Q(X)\mbox{,}
\end{eqnarray}
where $D$ is dimension of the target space. The  eqs.(22) define the
form of functions $A_{\alpha\beta}$ (4) in our case.  Ghost
contribution $G_{\alpha\beta}$ has the standard form as in the free
string theory.

As a result the first of eqs.(11) in our theory
gives standard conditions $\alpha_0 =\beta_0=1$,  $D=26$ and
new conditions
\begin{equation}
(\Box +m^2 )Q(X)=0
\end{equation}
\begin{equation}
\gamma^\prime (\tau , \sigma )=0
\end{equation}
where $m^2$ is the mass square of tachyon string mode. The
equation $\partial U^{\gamma}_{\alpha\beta}/\partial \tau$ is
fulfilled automatically in our case. The second of eqs.(11) leads
to the following condition under the eqs.(23, 24)\footnote{If the
equations (23, 24) are fulfilled then $d\Omega /d\tau \sim
\dot{\gamma}$ because of special structure of the Hamiltonian
(16). By the way we have just the case when the functions $T_\alpha$
depend explicitly on time.}
\begin{equation}
\dot{\gamma} (\tau,\sigma )=0
\end{equation}
The eq.(23) is the known free equation of motion for the tachyon
string mode. The eqs.(24, 25) mean that the string world
sheet should be flat.  Thus the above quantization procedure will be
consistent if the world sheet is flat and the field $Q(X)$ in the
Lagrangian (13) satisfies the tachyon equation of motion. Note
that in this case the operators $:L_n:$ and $:\bar{L}_n:$ form
quantum Virasoro algebra.

\section{Summary}
We have suggested the prescription allowing in certain cases to
construct a quantum gauge formulation starting from non-gauge
classical theory. The crucial role in this formulation is
played by the eqs.(11) defining values which the theory parameters
should take to provide consistency of the formally used
BFV-procedure.

To illustrate this prescription we have considered the closed
bosonic string in tachyon background field.
We have found that the above prescription works very well
and allows to obtain correct free equation of motion for
background field corresponding to tachyon string mode.
Quantum Virasoro algebra on the flat world sheet takes place under
this equation in spite of absence of any constraints in the
initial classical theory.

The obtained results show that the above procedure can be considered
as a general method allowing to construct gauge invariant
quantum theory for
an initially non-gauge classical model. In particular we hope that
this method provides a possibility to derive non-linear
equations of motion for strings interacting with massless and massive
background fields in framework of canonical quantization.

\section*{Acknowledgments}
The authors are very grateful to I.V.Tyutin for discussion of
some aspects of the paper.  The paper was finished during the
visit of I.L.B. to Institute of Physics, Humboldt Berlin University.
He would like to express the gratitude to D.Lust, D.Ebert, H.Dorn,
G.Gardoso, C.Preitschopf, C.Schubert and M.Schmidt for their
hospitality, support and interesting discussions.  The work of I.L.B.
and V.D.P. was supported in part by ISF under the grant No~RI1300 and
RFBR, project No~94-02-03234.

\newcounter{appendix}
\setcounter{equation}{0}
\setcounter{appendix}{1}
\renewcommand{\theappendix}{\Alph{appendix}}
\section*{Appendix \theappendix}
\renewcommand{\theequation}{\theappendix \arabic{equation}}
Let us consider standard mode expansion for the operators
$X^{\mu}$ and $P^{\mu}$ (see e.g.~\cite{Sup_87, Scherk_75})
\begin{eqnarray}
X^{\mu} & = &
\frac{1}{\sqrt{2\pi}}x_0^{\mu}+\sqrt{2\pi\alpha^\prime}p_0^{\mu}\tau
\nonumber\\
& + & \frac{i\sqrt{\alpha^\prime}}{\sqrt{2}}\sum_{n\neq
0}\frac{1}{n} (\alpha^{\mu}_{n}e^{-\imath n(\tau - \sigma)}
+\bar{\alpha}^{\mu}_{n}e^{-\imath n(\tau + \sigma)}),
\nonumber\\
P^{\mu} & = &
\frac{1}{\sqrt{2\pi\alpha^\prime}}p_0^{\mu}
+\frac{1}{2\pi\sqrt{\alpha^\prime}\sqrt{2}}\sum_{n\neq0}
(\alpha^{\mu}_{n}e^{-\imath n(\tau - \sigma)}
+\bar{\alpha}^{\mu}_{n}e^{-\imath n(\tau + \sigma)})
\end{eqnarray}
where the operators of zero modes $x_0^{\mu}$,
$p_0^{\mu}$ and of the oscillating ones $\alpha^{\mu}_{n}$,
$\bar{\alpha}^{\mu}_{n}$ satisfy the following commutation
relations:
\begin{equation}
[x_0^{\mu}, p_0^{\nu}]=i\hbar \delta_{\nu}^{\mu},\;
[\alpha^{\mu}_{m}, \alpha^{\nu}_{n}]=[\bar{\alpha}^{\mu}_{m},
\bar{\alpha}^{\nu}_{n}]=\hbar m\delta_{m,-n}\eta^{\mu\nu}
\end{equation}

We denote an arbitrary ordered operator $A$ depending on $x_0^{\mu}$,
$p_0^{\mu}$, $\alpha^{\mu}_{n}$, $\bar{\alpha}^{\mu}_{n}$ as $O(A)$.
The most general form of $O(\alpha^{\mu}_{n}\alpha^{\nu}_{m})$,
$O(\bar{\alpha}^{\mu}_{n}\bar{\alpha}^{\nu}_{m})$,
$O(x_0^{\mu}p_0^{\nu})$, $O(p_0^{\nu}x_0^{\mu})$ can be written as
follows
\begin{eqnarray}
O(\alpha^{\mu}_{n}\alpha^{\nu}_{m}) & = &
(1-c_{nm})\alpha^{\mu}_{n}\alpha^{\nu}_{m} +
c_{nm}\alpha^{\nu}_{m}\alpha^{\mu}_{n},
\nonumber\\
O(\bar{\alpha}^{\mu}_{n}\bar{\alpha}^{\nu}_{m}) & = &
(1-\bar{c}_{nm})\bar{\alpha}^{\mu}_{n}\bar{\alpha}^{\nu}_{m} +
\bar{c}_{nm}\bar{\alpha}^{\nu}_{m}\bar{\alpha}^{\mu}_{n},
\nonumber \\
O(x^{\mu}_0 p_{\nu 0}) & = & (1-c_0 )x^{\mu}_0
p_{\nu 0} + c_0 p_{\nu 0}x^{\mu}_0, \nonumber\\
O(p_{\nu 0}x^{\mu}_0 ) & = & (1-\bar{c}_0 )p_{\nu 0}x^{\mu}_0 +
\bar{c}_0 x^{\mu}_0 p_{\nu 0}
\end{eqnarray}
where the parameters $c_{mn}$, $\bar{c}_{mn}$, $c_0$, $\bar{c}_0$
characterize a specific choice of ordering prescription.
The commutation relation (A2)
and the symmetries of eq.(A3) lead to the properties
\begin{equation}
1-c_{mn}=c_{nm}\equiv c_n
\delta_{n,-m},\; 1-\bar{c}_{mn}=\bar{c}_{nm}\equiv \bar{c}_n \delta_{n,-m},
\; 1-\bar{c}_0=c_0
\end{equation}
Remember that every specific choice of the quantities $c_n$,
$\bar{c}_n$, and $c_0$ leads to a strictly defined type of ordering
prescription. For example, in the string models the Weyl ordering for
zero modes and the Wick ordering for the oscillating ones is
usually used (see e.g.~\cite{BuchFr_91}). In this case
\begin{equation}
c_n =\bar{c}_n=\Theta (n),\; c_0 ={1\over 2}
\end{equation}
The eqs.(A3) define the most general form of ordering prescription for
zero and oscillating string modes.
(The ordering prescription given by (A5) will be called
 the normal ordering
and the normal form of an operator $A$
will be denoted as ${\cal N}(A)$.)

It is well known~\cite{Sup_87, Scherk_75} that in the case of free
string the following relations take place
\begin{eqnarray}
L_n^{(0)} & \rightarrow & O(L_n^{(0)})={\cal  N}
(L_n^{(0)})-\hbar\alpha_0\delta_{n,0},
\nonumber \\
\bar{L}_n^{(0)} & \rightarrow & O(\bar{L}_n^{(0)})={\cal
N}(\bar{L}_n^{(0)})-\hbar\beta_0\delta_{n,0}
\end{eqnarray}
It is natural to consider only those ordering prescriptions (A3)
which are consistent with the eqs.(A6) accepted in the free string
theory.  It means that the eqs.(A6) can be treated as some
restrictions on  arbitrary parameters $c_{mn}$, $\bar{c}_{mn}$ and
$c_0$ in the eqs.(A3,A4).  If we demand the eqs.(A6) to take
place and the parameters $\alpha_0$, $\beta_0$ to have  fixed
values we see that  $c_{mn}$, $\bar{c}_{mn}$, $c_0$ should depend
on the  parameters $\alpha_0$ and $\beta_0$. Straightforward
calculations with the use of the eqs.(A3 - A6) and the definitions of
$L^{(0)}$ and $\bar{L}^{(0)}$ lead to
\begin{equation}
\alpha_0=-{D\over2}\sum_{n>0}n(1-c_n +c_{-n}),\; \beta_0=-{D\over
2}\sum_{n>0}n(1-\bar{c}_n +\bar{c}_{-n})
\end{equation}
A solution (not general but nevertheless acceptable for our aims)
of these equations looks as follows
\begin{eqnarray}
& & c_0 =1/2,\; 1-c_{n}=c_{-n}=-\alpha_0\frac{(1-\mu )^2}{\mu D}\mu^n
\nonumber \\
& & 1-\bar{c}_{n}=\bar{c}_{-n}=-\beta_0\frac{(1-\mu
)^2}{\mu D}\mu^n,\; n>0,\; |\mu|<1
\end{eqnarray}
where $\mu$ is an  arbitrary parameter. We suppose as usually that
$c_0={1\over 2}$.  As a result all the coefficients $c_{mn}$,
$\bar{c}_{mn}$, $c_0$ and $\bar{c}_0$ are defined now in terms of
the given parameters $\alpha_0$, $\beta_0$ and an additional
parameter $\mu$. Using the obtained coefficients $c_{mn}$,
$\bar{c}_{mn}$, $c_0$ and $\bar{c}_0$ we find
\begin{eqnarray}
:\alpha^{\mu}_{n}\alpha^{\nu}_{m}: & = &
(1+\alpha_0 \frac{(1-\mu )^2}{\mu D}\mu^n \delta_{n,-m})
\alpha^{\mu}_{n}\alpha^{\nu}_{m} -
\alpha_0 \frac{(1-\mu )^2}{\mu D}\mu^n
\delta_{n,-m}\alpha^{\nu}_{m}\alpha^{\mu}_{n},
\nonumber \\
:\bar{\alpha}^{\mu}_{n}\bar{\alpha}^{\nu}_{m}: & = &
(1+\beta_0 \frac{(1-\mu )^2}{\mu D}\mu^n \delta_{n,-m})
\bar{\alpha}^{\mu}_{n}\bar{\alpha}^{\nu}_{m} -
\beta_0 \frac{(1-\mu )^2}{\mu D}\mu^n \delta_{n,-m}
\bar{\alpha}^{\nu}_{m}\bar{\alpha}^{\mu}_{n},
\nonumber \\
:x^{\mu}_0 p_{\nu 0}: & = & {1\over 2}
(x^{\mu}_0 p_{\nu 0} +  p_{\nu 0}x^{\mu}_0)
\end{eqnarray}
Here and further  we apply the notation $:A:$ for any operator
$A$ ordered according to the prescription (A9).

\setcounter{equation}{0}
\setcounter{appendix}{2}
\section*{Appendix \theappendix}

A symbol of an operator is a $c-$valued function of phase variables
corresponding to some operator ordering. Let $A$
is the symbol of operator $\hat{A}$, $B$ is the symbol of operator
$\hat{B}$. Then the symbol corresponding to operator $\hat{A}\hat{B}$
is denoted  $A*B$ and looks like this~\cite{Ber_83, Vas_76}:
\begin{equation}
\stackrel{\displaystyle A*B=\exp(\Gamma_1^M \Gamma_2^N
\frac{\delta}{\delta\Gamma_1^M} \frac{\delta}{\delta\Gamma_2^N})
A(\Gamma_1)B(\Gamma_2)
\biggr|_{\displaystyle \Gamma_1 =\Gamma_2 =\Gamma} }
{\rule{0.4pt}{2mm}\rule{6mm}{0.4pt}\rule{0.4pt}{2mm}
\hspace{43mm}}
\end{equation}
where
\begin{equation}
\stackrel{\displaystyle { }\Gamma_1^M \Gamma_2^N
=\Gamma_1^M \Gamma_2^N -:\Gamma_1^M \Gamma_2^N:}
{\rule{0.4pt}{2mm}\rule{6mm}{0.4pt}
\rule{0.4pt}{2mm}\hspace{41mm}}
\end{equation}
are fundamental contractions of the canonical variables
$\Gamma^M$. The symbol corresponding to commutator of operators
has the form
\begin{equation}
[A, B]_* =A*B-B*A
\end{equation}

 We apply formalism of operators symbols to
calculate quantum algebra of constraints~(21).  Phase space
variables $\Gamma^M$ in our theory are:
\begin{equation}
\Gamma^M=(X^{\mu}(\tau ,z), P^{\mu}(\tau ,z))
\end{equation}
where $z\equiv e^{\imath\sigma}$ and we use the ordering prescription
given in the Appendix A.

According to the eqs.(B3,B1)  $*-$commutator of symbols
has the following general structure: $[A,B]_{*} = i\hbar\{A,B\} +
\Delta_{[A,B]}^{(2)} + O(\hbar^{3})$ where $\{A,B\}$ is  Poisson
bracket and $\Delta_{[A,B]}^{(2)}$ is proportional to $\hbar^{2}$.
The straightforward but tedious enough calculations lead to the
following explicit form of $\Delta_{[A,,B]}^{(2)}$:
\begin{eqnarray}
 & & \stackrel{\displaystyle \Delta_{[A,
B]}^{(2)}  =  \oint\frac{dz_1}{iz_1}\oint\frac{dz_2}{iz_2}
\bigl\{ (
[X^\mu_1 (\tau ,z_1)P^\nu_1 (\tau ,z_2)]_A }
{\hspace{29mm}\rule{0.4pt}{2mm}\rule{16mm}{0.4pt}
\rule{0.4pt}{2mm}}
\nonumber \\
 & & \stackrel{\displaystyle \times
(X^\mu_2 (\tau ,z_1) X^\nu_2(\tau ,z_2)
\frac{\partial^2 A}{\partial X^\mu_1\partial X^\mu_2}(\tau ,z_1)
\frac{\partial^2 B}{\partial P^\nu_1\partial X^\nu_2}(\tau ,z_2) }
{\rule{0.4pt}{2mm}\rule{16mm}{0.4pt}
\rule{0.4pt}{2mm}\hspace{65mm}}
\nonumber \\
 & & \stackrel{\displaystyle +
X^{\prime\mu}_2 (\tau ,z_1) X^\nu_2(\tau ,z_2)
\frac{\partial^2 A}{\partial X^{\mu}_1\partial
X^{\prime\mu}_2}(\tau ,z_1)
\frac{\partial^2 B}{\partial P^\nu_1\partial X^\nu_2}(\tau ,z_2)}
{\rule{0.4pt}{2mm}\rule{17mm}{0.4pt}
\rule{0.4pt}{2mm}\hspace{63mm}}
\nonumber \\
 & & \stackrel{\displaystyle +
X^\mu_2 (\tau ,z_1) X^{\prime\nu}_2(\tau ,z_2)
\frac{\partial^2 A}{\partial X^\mu_1\partial X^\mu_2}(\tau ,z_1)
\frac{\partial^2 B}{\partial P^\nu_1\partial
X^{\prime\nu}_2}(\tau ,z_2)}
{\rule{0.4pt}{2mm}\rule{16mm}{0.4pt}
\rule{0.4pt}{2mm}\hspace{66mm}}
\nonumber \\
 & & \stackrel{\displaystyle +
P^\mu_2 (\tau ,z_1) P^\nu_2(\tau ,z_2)
\frac{\partial^2 A}{\partial X^\mu_1\partial P^\mu_2}(\tau ,z_1)
\frac{\partial^2 B}{\partial P^\nu_1\partial P^\nu_2}(\tau ,z_2)}
{\rule{0.4pt}{2mm}\rule{16mm}{0.4pt}
\rule{0.4pt}{2mm}\hspace{64mm}}
\nonumber \\
 & & \stackrel{\displaystyle +
X^{\prime\mu}_2 (\tau ,z_1) X^{\prime\nu}_2(\tau ,z_2)
\frac{\partial^2 A}{\partial X^\mu_1\partial
X^{\prime\mu}_2}(\tau ,z_1)
\frac{\partial^2 B}{\partial P^\nu_1\partial
X^{\prime\nu}_2}(\tau ,z_2))}
{\rule{0.4pt}{2mm}\rule{17mm}{0.4pt}
\rule{0.4pt}{2mm}\hspace{67mm}}
\nonumber \\
 & & \stackrel{\displaystyle +
[X^{\prime\mu}_1 (\tau ,z_1)P^\nu_1 (\tau ,z_2)]_A }
{\rule{0.4pt}{2mm}\rule{17mm}{0.4pt}
\rule{0.4pt}{2mm}\hspace{10mm}}
\nonumber \\
 & & \stackrel{\displaystyle \times
(X^\mu_2 (\tau ,z_1) X^\nu_2(\tau ,z_2)
\frac{\partial^2 A}{\partial X^{\prime\mu}_1\partial
X^\mu_2}(\tau ,z_1)
\frac{\partial^2 B}{\partial P^\nu_1\partial X^\nu_2}(\tau ,z_2)}
{\rule{0.4pt}{2mm}\rule{16mm}{0.4pt}
\rule{0.4pt}{2mm}\hspace{64mm}}
\nonumber \\
 & & \stackrel{\displaystyle +
 X^{\prime\mu}_2 (\tau ,z_1) X^\nu_2(\tau ,z_2)
\frac{\partial^2 A}{\partial X^{\prime\mu}_1\partial
X^{\prime\mu}_2}(\tau ,z_1)
\frac{\partial^2 B}{\partial P^\nu_1\partial X^\nu_2}(\tau ,z_2)}
{\rule{0.4pt}{2mm}\rule{17mm}{0.4pt}
\rule{0.4pt}{2mm}\hspace{65mm}}
\nonumber \\
 & & \stackrel{\displaystyle +
X^\mu_2 (\tau ,z_1) X^{\prime\nu}_2(\tau ,z_2)
\frac{\partial^2 A}{\partial X^{\prime\mu}_1\partial
X^\mu_2}(\tau ,z_1)
\frac{\partial^2 B}{\partial P^\nu_1\partial
X^{\prime\nu}_2}(\tau ,z_2)}
{\rule{0.4pt}{2mm}\rule{16mm}{0.4pt}
\rule{0.4pt}{2mm}\hspace{65mm}}
\nonumber \\
 & & \stackrel{\displaystyle +
X^{\prime\mu}_2 (\tau ,z_1) X^{\prime\nu}_2(\tau ,z_2)
\frac{\partial^2 A}{\partial X^{\prime\mu}_1\partial
X^{\prime\mu}_2}(\tau ,z_1)
\frac{\partial^2 B}{\partial P^\nu_1\partial
X^{\prime\nu}_2}(\tau ,z_2))}
{\rule{0.4pt}{2mm}\rule{17mm}{0.4pt}
\rule{0.4pt}{2mm}\hspace{67mm}}
\nonumber \\
 & & \stackrel{\displaystyle +
P^\mu_2 (\tau ,z_1) P^\nu_2(\tau ,z_2)
\frac{\partial^2 A}{\partial X^{\prime\mu}_1 \partial
P^\mu_2}(\tau ,z_1)
\frac{\partial^2 B}{\partial P^\nu_1 \partial P^\nu_2}(\tau ,z_2)
))  }
{\rule{0.4pt}{2mm}\rule{16mm}{0.4pt}
\rule{0.4pt}{2mm}\hspace{65mm}}
\nonumber \\
 &  & -( A\longleftrightarrow B)\bigr\}
\end{eqnarray}
where subscript index $A$ denotes antisymmetric combination of the
contractions in the braces and $X^{\prime\nu}(\tau ,z)\equiv
iz\frac{d}{dz}X^{\nu}(\tau ,z)$. In the case under consideration the
constraints are quadratic in momenta. It means that the
terms corresponding to $\hbar^{n}$, $n > 2$ are absent in the
$*-$commutator and $\Delta_{[A,B]}^{(2)}$ is the only quantum
contribution.

Taking into account the ordering prescription given in Appendix A
we find after straightforward calculations
\begin{eqnarray}
\lefteqn{\stackrel{\displaystyle X^\mu (\tau ,z_1) X^\nu (\tau ,z_2)}
{\rule{0.4pt}{2mm}\rule{16mm}{0.4pt}
\rule{0.4pt}{2mm}\hspace{12mm}}}
\nonumber \\
 & = & \frac{\alpha{^\prime}\hbar}{2}\eta^{\mu\nu}
\sum_{n>0}{1\over n}[(z_1/z_2)^n +(z_2/z_1)^n]
+ i\hbar\tau\eta^{\mu\nu}
\nonumber \\
& + & \frac{\alpha{^\prime}\hbar}{2}(\alpha_0 +\beta_ 0)
\frac{(1-\mu )^2 }{\mu D}\eta^{\mu\nu}
\sum_{n>0}{1\over n}[(\mu z_1/z_2)^n +(\mu z_2/z_1)^n],
\nonumber \\
\lefteqn{\stackrel{\displaystyle X^\mu (\tau ,z_1) P^\nu (\tau ,z_2)}
{\rule{0.4pt}{2mm}\rule{16mm}{0.4pt}
\rule{0.4pt}{2mm}\hspace{12mm}}}
\nonumber \\
 & = & {1\over 2}\frac{i\hbar }{2\pi }\eta^{\mu\nu}
(1+\sum_{n>0}[(z_1/z_2)^n +(z_2/z_1)^n])
\nonumber \\
& + & {1\over 2}\frac{i\hbar }{2\pi }(\alpha_0 -\beta_ 0)
\frac{(1-\mu )^2 }{\mu D}\eta^{\mu\nu}
\sum_{n>0}{1\over n}[(\mu z_1/z_2)^n +(\mu z_2/z_1)^n],
\nonumber \\
\lefteqn{\stackrel{\displaystyle P^\mu (\tau ,z_1) P^\nu (\tau ,z_2)}
{\rule{0.4pt}{2mm}\rule{16mm}{0.4pt}
\rule{0.4pt}{2mm}\hspace{12mm}}}
\nonumber \\
 & = & {1\over 2}\frac{\hbar }{(2\pi )^2 \alpha{^\prime}}\eta^{\mu\nu}
\sum_{n>0}n[(z_1/z_2)^n +(z_2/z_1)^n]
\nonumber \\
& + &{1\over 2}\frac{\hbar }{(2\pi )^2 \alpha{^\prime}}
(\alpha_0 +\beta_ 0)\frac{(1-\mu )^2 }{\mu D}\eta^{\mu\nu}
\sum_{n>0}n[(\mu z_1/z_2)^n +(\mu z_2/z_1)^n]
\end{eqnarray}
Since $|\mu|<1$ and $|z_1|=|z_2|=1$ all power series in the variables
$\mu z_1 /z_2 $ and $\mu z_2/z_1 $ in the eqs.(B6)
are convergent. However, power series in the variables
$z_1 /z_2 $ and $z_2/z_1 $  diverge.\footnote{Note  that
divergent series are absent in the case of normal ordering
contractions (see ref.~\cite{BuchFr_91}).} Moreover, the singular
point $z_1=z_2$ in the divergent series is situated on the contour of
integration. Hence the corresponding integrals are ill defined and a
regularization is required.

To regularize the integrals and contractions we change $z_1 /z_2$  by
$(z_1 /z_2 )e^{-\epsilon}$ and $z_2 /z_1$  by $(z_2
/z_1)e^{-\epsilon}$ in the divergent series. Here $\epsilon>0$ is the
regularization parameter. As a result all series in the eqs.(B6) are
summed to elementary functions and contractions take the form:
\begin{eqnarray}
\lefteqn{\stackrel{\displaystyle X^\mu (\tau ,z_1) X^\nu (\tau ,z_2)}
{\rule{0.4pt}{2mm}\rule{16mm}{0.4pt}
\rule{0.4pt}{2mm}\hspace{12mm}}}
\nonumber \\
 & = & \frac{\alpha{^\prime}\hbar}{2}\eta^{\mu\nu}
[{\rm ln}(1-z_1 e^{-\epsilon}/z_2 )+{\rm ln}(1-z_2 /(e^{\epsilon}z_1 ))]
+ i\hbar\tau\eta^{\mu\nu}
\nonumber \\
& + &\frac{\alpha{^\prime}\hbar}{2}(\alpha_0 +\beta_ 0)
\frac{(1-\mu )^2 }{\mu D}\eta^{\mu\nu}
[{\rm ln}(1-\mu z_1/z_2 )+{\rm ln}(1-\mu z_2 /z_1 )],
\nonumber \\
\lefteqn{\stackrel{\displaystyle X^\mu (\tau ,z_1) P^\nu (\tau ,z_2)}
{\rule{0.4pt}{2mm}\rule{16mm}{0.4pt}
\rule{0.4pt}{2mm}\hspace{12mm}}}
\nonumber \\
 & = & {1\over 2}\frac{i\hbar }{2\pi }\eta^{\mu\nu}
\biggl (\frac{z_1 e^{-\epsilon}}{z_2 -z_1 e^{-\epsilon}}
-\frac{z_1 e^{\epsilon}}{z_2 -z_1 e^{\epsilon}}\biggr )
\nonumber \\
& + & {1\over 2}\frac{i\hbar }{2\pi }(\alpha_0 -\beta_ 0)
\frac{(1-\mu )^2 }{\mu D}\eta^{\mu\nu}
\biggl (\frac{\mu z_1 }{z_2 -\mu z_1 }+\frac{\mu ^{-1}z_1 }{z_2 -\mu ^{-1}z_1 }
\biggr )
\nonumber \\
\lefteqn{\stackrel{\displaystyle P^\mu (\tau ,z_1) P^\nu (\tau ,z_2)}
{\rule{0.4pt}{2mm}\rule{16mm}{0.4pt}
\rule{0.4pt}{2mm}\hspace{12mm}}}
\nonumber \\
 & = & {1\over 2}\frac{\hbar }{(2\pi )^2 \alpha{^\prime}}\eta^{\mu\nu}
\biggl (\frac{z_2 z_1 e^{-\epsilon}}{(z_2 -z_1 e^{-\epsilon})^2 }
-\frac{z_2 z_1 e^{\epsilon}}{(z_2 -z_1 e^{\epsilon})^2 }\biggr )
\nonumber \\
& + &{1\over 2}\frac{\hbar (\alpha_0 +\beta_ 0)}{(2\pi )^2 \alpha{^\prime}}
\frac{(1-\mu )^2 }{\mu D}\eta^{\mu\nu}
\biggl (\frac{\mu z_2 z_1 }{(z_2 -\mu z_1 )^2 }
+\frac{\mu ^{-1}z_2 z_1 }{(z_2 -\mu ^{-1}z_1 )^2 }\biggr )
\end{eqnarray}
Now the integrals in the eqs.(B5) can be calculated  by standard
methods and  at the end of calculations we
should set $\epsilon\rightarrow 0 $.

We will describe briefly the procedure of calculation of the
integrals.  First, integrate over the variable $z_2$.  As it
has been noted  the regularization leads to the situation when two
singular points $z_2=z_1 e^{-\epsilon}$  and $z_2=z_1 e^{\epsilon}$
arising in the integrand turned out to be situated on the opposite
sides of contour.  The only pole in the case under consideration  is
$z_2=z_1 e^{-\epsilon}$.
Thus, this regularization procedure can in principle be treated as
a some sort of point splitting adapted for use within the
canonical formalism.

The above procedure leads, for example, to the following result
\begin{eqnarray}
\lefteqn{ \stackrel{\displaystyle
[X^\mu (\tau ,z_1)P^\nu (\tau ,z_2)]_A  }
{\rule{0.4pt}{2mm}\rule{16mm}{0.4pt}
\rule{0.4pt}{2mm}\hspace{13mm}}}
\nonumber \\
 & = & {1\over 2}\frac{i\hbar }{2\pi }\eta^{\mu\nu}
\biggl (\frac{z_1 e^{-\epsilon}}{z_2 -z_1 e^{-\epsilon}}
-\frac{z_1 e^{\epsilon}}{z_2 -z_1 e^{\epsilon}}\biggr )
\end{eqnarray}
Since the eq.(B8) or its derivative with respect to $z_1$ is contained
in  all $*-$com\-mu\-ta\-tors (see (B3)) residues in  all the
poles\footnote{It is evident that points $z_2=\mu z_1 $ and $z_2=0$
are poles whereas the other singular points
$z_2=\mu ^{-1}z_1 $  and $z_2=z_1 e^{\epsilon}$ are not.}
except $z_2=z_1 e^{-\epsilon}$ will vanish if
$\epsilon\rightarrow 0 $. The method of calculation of the integrals
within this  regularization procedure  is consistent with
the method described  in ref.~\cite{Scherk_75} and used then
in ref.~\cite{BuchFr_91} where  boundary of the ring
 $(|z_2|>|z_1|)-(|z_1|>|z_2|)$
played role of a contour for non-regularized integrals. In order
to compute $*-$com\-mu\-ta\-tors of the constraints
symbols (21) we should replace $A$ and $B$ in the eq.(B5) by the $L_n$
or the $\bar{L}_n$ and then integrate the obtained expressions
according to the above prescription.

\end{document}